\documentclass[twocolumn]{aastex631}
\usepackage{url}
\usepackage{amsmath}
\usepackage{placeins}
\shorttitle{Collapsed SMDSs as late stage quasi-stars}
\shortauthors{Ilie}

\begin{document}
\title{JWST's Little Red Dots as collapsed Supermassive Dark Stars}

\author{Cosmin Ilie}
\affiliation{Colgate University\\
13 Oak Drive\\
Hamilton, NY 13346}

\correspondingauthor{Cosmin Ilie}
\email{cilie@colgate.edu}


\begin{abstract}
The nature of the ``Little Red Dots'' (LRDs) is one of the most profound mysteries posed by the JWST data.
One promising class of models that can reproduce the observed LRDs spectra and morphology are quasi-stars: massive envelopes surrounding accreting black holes formed via the collapse of supermassive stars (SMSs).
However, the canonical SMS pathway relies on a highly restricted set of environmental and structural conditions: strong Lyman--Werner (LW) backgrounds to suppress H$_2$ cooling, high and sustained gas inflow rates to enforce entropy stratified envelopes, and assume non-zero rotational support in order to prevent GR instability collapse before $\sim 10^6 M_{\odot}$.
Here we show that supermassive dark stars (SMDSs), powered by dark matter (DM) annihilation rather than nuclear burning, naturally satisfy the key structural and energetic requirements for quasi-star (QS) formation while relaxing {\it all} of those restrictive conditions listed above. Moreover, quasi-stars formed through the SMDS pathway are born with prompt BH masses
\(\gtrsim 10\%\) of the progenitor mass. They therefore enter directly into a
late-stage quasi-star regime; subsequently the envelope expands and cools until its photosphere reaches the zero-metallicity opacity limit \(T_{\rm eff}\sim3000\)--\(6000\,{\rm K}\). Those cool, optically thick, unresolved photospheres can reproduce key features of many JWST LRDs. 
\end{abstract}


\section{Introduction}\label{sec:intro}

JWST has revealed a population of compact, extremely red, and highly luminous sources at $z \gtrsim 7$ that are commonly referred to as ``Little Red Dots'' (LRDs; e.g., \citealt{Kocevski2024,Wang2024,LiInayoshi2025,Taylor2025,VaidaFarber2026}).
Their spectral energy distributions (SEDs) and inferred luminosities, $L \sim 10^{11}$--$10^{12}\,L_\odot$, challenge standard stellar~\citep{Baggen2024densities,Setton2024vshape} or AGN interpretations~\citep{Akins2025cosmos,Durodola2025agn,Trinca2024superedd}.
Several works have argued that some LRDs may host rapidly growing black holes (BHs) embedded in optically thick, dusty or gaseous flows, producing cool, extended photospheres at $T_\mathrm{eff} \sim 3000$--$6000\,$ \citep[e.g.,][]{Wang2024,LiInayoshi2025,DeGraaff2025,Naidu2025,VaidaFarber2026}.\footnote{Sometimes those configurations are called BH-stars in the literature~\citep{Naidu2025}.}
As shown by \cite{Begelman_2025}, one physically motivated manifestation of such systems is the quasi-star: an accreting BH embedded in a massive, radiation-supported, Compton thick envelope~\citep{BegelmanRossiArmitage2008,Begelman2010,VolonteriBegelman2010}.

In the canonical picture, quasi-stars form when a supermassive star (SMS) collapses via GR instability, leaving behind a small seed BH in hydrostatic equilibrium with the remaining envelope.
The BH accretes at or near the Eddington limit of the \emph{total} mass (envelope plus BH), while the envelope readjusts to a convective, radiation-pressure-supported configuration.
The outer layers expand towards a Hayashi-like limit set by the composition and opacity, yielding a cool, extended photosphere that can in principle reproduce LRD-like colors and luminosities \citep{Hassan2025,Begelman_2025}.

However, the SMS-based quasi-star pathway is subject to a number of stringent requirements.
First, SMS formation at $M_\star \gtrsim 10^5$--$10^6\,M_\odot$ in primordial gas generally requires strong LW radiation backgrounds to suppress H$_2$ cooling and keep gas in the atomic-cooling regime, avoiding fragmentation into normal Population III stars \citep{Begelman2006,Dijkstra2008,Latif2013,Haiman2013}.
Second, rapid inflow rates $\dot{M}_\mathrm{gas} \gtrsim 0.1$--$1\,M_\odot\,\mathrm{yr}^{-1}$ are needed to build the SMS quickly and maintain a non-relaxed, hylotropic envelope whose entropy increases with enclosed mass \citep{Begelman2010,Woods2017}.
Third, a modest amount of rotation is required to stabilise the SMS against early GR collapse; non-rotating SMSs become unstable at masses of only a few $\times 10^5\,M_\odot$, well below the canonical $10^6\,M_\odot$ scale \citep{FullerWoosleyWeaver1986ApJ...307..675F,BaumgarteShapiro1999,Haemmerle2021,Hori2023}.
Finally, the classical quasi-star solutions themselves typically allow BH-to-envelope mass fractions of only $M_\mathrm{BH}/M_\star \lesssim 10^{-2}$ before the photosphere reaches the Hayashi limit and the envelope disperses \citep{BegelmanRossiArmitage2008,VolonteriBegelman2010}. Recent saturated-convection models show that quasi-star envelopes may persist to much larger BH fractions~\citep{Coughlin_Begelman_2024}, making ``late-stage'' quasi-stars (i.e. $M_{BH}\gtrsim 0.1 M_{SMS}$) especially compelling LRD analogues~\citep{Begelman_2025}.  

Supermassive dark stars (SMDSs) offer a qualitatively different route to similar objects. SMDSs are zero metallicity gas clouds in hydrostatic equilibrium, whose luminosity is powered predominantly by DM annihilation rather than nuclear burning \citep{Spolyar2008,Freese2010,Ilie2012,Freese2016}.
Because DM heating supplies energy throughout the interior, SMDSs can remain cool ($T_{eff}\lesssim$ few $\times 10^{4}$~K) and extended ($R\sim A.U.$s) while accreting at comparatively modest baryonic rates ($\dot{M}\sim10^{-2} M_\odot$/yr), and can reach $M_\star \sim 10^6 M_\odot$ without ever igniting conventional main-sequence nuclear burning. As shown in \cite{Ilie:2025zzj}, SMDSs BH remnants could provide a solution to the puzzle posed by origin of the Supermassive Black Holes powering the most distant quasars observed, such as UHZ1. Moreover, mergers of SMBHs seeded by Dark Stars could contribute significantly to the observed PTA gravitational background wave signal~\citep{ghodla2025reconstructingptameasurementsearly}. Lastly, several photometric and spectroscopic supermassive dark star candidates have alraedy been identified in the JWST data~\citep{Ilie_2023,ilie2025spectroscopic}.


In this paper we argue that SMDSs are natural progenitors of LRDs, a point we first
hypothesized in \cite{Ilie:2025zzj}.
Here we explicitly show that SMDSs satisfy the structural and energetic conditions for GR collapse and bound-envelope survival that appear in the SMS literature.
A key difference with respect to the classical quasi-star scenario is that in the SMDS pathway proposed here, the mass of the promptly formed BH can be a significant part of the mass of the progenitor star. As such, Black Hole star configurations generated by collapsed SMDSs are akin to late-stage quasi-stars, i.e. configurations where the central BH mass is greater than $\sim 10\%$ of the mass of the progenitor SMS.
Moreover we show that the envelope binding energy, prompt collapse energy, and subsequent time-integrated accretion feedback place the system in a ``Goldilocks'' window: the envelope can remain bound while still being energetically susceptible to inflation into a cool, extended, quasi-star-like configuration.


The paper is organised as follows.
Section~\ref{sec:smds_collapse} discusses the onset of general relativistic (GR) instabilities in dark stars, summarizing the main results of \cite{FreeseIlie2025}. In the same section our choice for a fiducial estimate of mass of the promptly formed BH is motivated.
Section~\ref{sec:qs_formation} describes the possible formation of a quasi-star-like remnant from the SMDS collapse product, compares the SMDS and SMS pathways, and highlights the structural and environmental advantages of the SMDS channel.
Section~\ref{sec:lrd_connection} connects SMDS-born quasi-star-like remnants to JWST LRDs, focusing on the inflation and cooling of the outer envelope and checking the requirement of maintaining deep Compton thickness in order to obscure the emission from the embedded BH. 
We conclude in Section~\ref{sec:conclusions}.

\section{Collapse of Supermassive Dark Stars}\label{sec:smds_collapse}

The collapse of SMDSs was first analyzed in \cite{FreeseIlie2025}, where accreting dark stars were followed until the onset of the Feynman--Chandrasekhar general-relativistic instability. The instability endpoint was identified when the evolved central density exceeded the critical value for GR radial instability. We adopt as our fiducial pre-collapse model a non-rotating SMDS powered by 100 GeV WIMP annihilation at this central-density-selected onset. Based on the analysis of \cite{FreeseIlie2025}, this occurs at $M_\star \simeq 2.6\times10^6\,M_\odot$. In Sec.~\ref{ssec:StructProps} we list the model's basic structural properties, then check the onset of GR instability using the first adiabatic exponent criterion for radiation pressure supported stars. In Sec.~\ref{ssec:PromtBHEstimate} we motivate a fiducial prompt black-hole mass using the structure of the \texttt{MESA} profile and a local GR-stiffness scale estimate, whereas in Sec.~\ref{ssec:BindingEnergy} we show that the extended envelope is likely to remain bound. 

\subsection{Structural properties at GR onset}\label{ssec:StructProps}

The SMDS model at GR onset exhibits several key properties, summarized below. It has a total mass $M_\star \simeq 2.6\times10^6\,M_\odot$ and photospheric radius $R_\star \simeq 1.7\times10^4\,R_\odot$. Its effective temperature $T_\mathrm{eff} \simeq 2.6\times10^4\,$K. Its luminosity is near Eddington: $L \simeq L_{\rm Edd}(M_\star)\simeq 1.0\times10^{11}\,L_\odot$. Moreover the internal structure has the following key properties:

\begin{itemize}
    \item Strong radiation domination in the interior, with $P_\mathrm{gas}/P_\mathrm{tot} \lesssim 10^{-2}$ over most of the enclosed mass. As such, the $n=3$ polytrope is a reasonable approximation for most of the star.~\footnote{In the standard SMS channel, rapid accretion produces a hylotropic structure: a compact, nuclear-burning, radiation-pressure-dominated core embedded in an extended, entropy-stratified envelope \citep[e.g.][]{Begelman2010}.} 
    \item first adiabatic exponent $\Gamma_1 \equiv (\partial\ln P/\partial\ln\rho)_s \approx 4/3$ to within $\sim 10^{-3}$, suggesting a star on the verge of collapse.
    \item Significant convective energy transport in our \texttt{MESA} models: the Schwarzschild criterion indicates instability to convection in the entire star.
\end{itemize}

The stability of radiation-dominated, $n=3$-like configurations against general-relativistic (GR) collapse can be characterized by a GR-corrected critical adiabatic index $\Gamma_{\rm crit}$, which modifies the Newtonian condition $\Gamma_1 > 4/3$ to $\Gamma_1 > \Gamma_{\rm crit}$, where
\begin{equation}
\Gamma_{\rm crit} \simeq \frac{4}{3} + C\,\frac{G M}{R c^2},
\end{equation}
with $C \sim 2$--3 for $n=3$ polytropes \citep[e.g.,][]{Chandrasekhar1964,ShapiroTeukolsky1983}. In this work, we implement it as a \emph{pressure-weighted} condition by computing the pressure-averaged adiabatic index
\begin{equation}
\langle \Gamma_1 \rangle_P \equiv \frac{\int \Gamma_1 P\,{\rm d}V}{\int P\,{\rm d}V}
\end{equation}
from the \texttt{MESA} models along the SMDS sequence, and comparing it to $\Gamma_{\rm crit}$ evaluated using the corresponding compactness $G M / (R c^2)$. For the SMDS configuration that we identify as being at the onset of GR instability, we find $\langle \Gamma_1 \rangle_P \approx 1.334$, essentially equal to $\Gamma_{\rm crit}$ for reasonable choices $C=2$--3, confirming that this model lies at the threshold of GR marginal stability. For a slightly more massive, otherwise similar configuration along the same sequence, the pressure-weighted average drops to $\langle \Gamma_1 \rangle_P \approx 1.33395$, which is below $\Gamma_{\rm crit}$ for both $C=2$ and $C=3$; this shows that the sequence has crossed into the GR-unstable regime and independently confirms that our fiducial SMDS progenitor is poised at the brink of collapse, with further growth driving it into global GR instability.

\subsection{Fiducial prompt black-hole mass}
\label{ssec:PromtBHEstimate}
\label{ssec:prompt_bh_mass}

The prompt black-hole mass expected from SMDS collapse is expected to be significantly larger than that of a standard SMS quasi-star progenitor.  The later scenario relies on
rapidly accreting SMSs which are hylotropic objects: a relatively small,
low-entropy convective core is embedded inside a massive, high-entropy envelope
\citep{Begelman2010,Woods2017,Haemmerle2021}.  The onset of GR instability is
then naturally associated with collapse of this core, so the initial black-hole
seed can be only a few percent of the progenitor mass. A larger prompt BH would prematurely push the system toward the terminal, Hayashi-limited regime. This is one of the reasons for which the SMS quasi-star formation pathway relies high sustained accretion rates ($\dot{M}\gtrsim 0.1 M_\odot/yr$): in order to maintain a very massive hylotropic envelope which will not collapse once the core collapsed.

In the SMDS pathway, by contrast, the progenitor is already DM-inflated and weakly bound, so even a large prompt BH could remain embedded in a massive optically thick envelope, as we will discuss in detail later in the manuscript. SMDSs  are extended,
radiation-pressure-dominated configurations much closer to globally relaxed
\(n\simeq3\)-like stars than to strongly core-envelope-separated hylotropes. 
We therefore expect the prompt black hole produced by SMDS collapse to be a
substantially larger fraction of the star than in the classical SMS
quasi-star channel.

The prompt mass is nevertheless not fixed uniquely by the onset of the GR
instability.  It depends on the nonlinear collapse: when an apparent horizon
first forms, how rapidly neighboring shells accrete through it, and whether
pressure waves, shocks, radiation transport, or fallback delay the collapse of
the outer layers.  In the absence of a full GR-hydrodynamic calculation, we
adopt a fiducial value guided by the structure of the \texttt{MESA} model near
the instability endpoint.

To estimate the mass of the prompt BH produced by the collapse our fiducial SMDS  we use  the following local stiffness condition: $\Gamma_1(r)> \Gamma_{\rm crtit}(r)$, with 
\begin{equation}
\Gamma_{\rm crit}(r)
=
\frac{4}{3}
+
C\,\frac{P(r)}{\rho(r)c^2},
\qquad C\simeq2\text{--}3 .
\label{eq:local_gamma_crit_pressure}
\end{equation}
This is not a local substitute for a full radial stability or collapse
calculation; the GR instability is fundamentally global.  Rather, it indicates
which part of the profile is closest to losing pressure support.  Applied the
the near-onset SMDS \texttt{MESA} sequence considered here, the \(C=3\) choice identifies
a central unstable region of order a few \(10^5\,M_\odot\), growing to
\(\sim6\times10^5\,M_\odot\) in the nearest post-threshold saved profile.  The
\(C=2\) choice is more conservative and delays the appearance of such a
region.  Thus the estimate is coefficient-dependent, but it robustly indicates
that the dynamically relevant inner mass scale is not a tiny few-percent core.
It is already several \(10^5\,M_\odot\), and plausibly approaches
\(10^6\,M_\odot\) once neighboring shells participate in the nonlinear collapse.

We therefore adopt $M_{\rm BH,0}\simeq 10^6\,M_\odot$
as a fiducial prompt black-hole mass.  This value should be understood as a
physically motivated scale, not as a sharply derived mass cut.  It is
consistent with the local GR-stiffness estimate, is substantially larger than
the few-percent seeds characteristic of hylotropic SMS quasi-star models, and
captures the expectation that SMDS collapse is less core-localized.  For the
this choice leaves
\begin{equation}
M_{\rm env}
\simeq
M_\star-M_{\rm BH,0}
\simeq
1.6\times10^6\,M_\odot .
\label{eq:fiducial_envelope_mass}
\end{equation}

We do not claim that the remaining envelope necessarily avoids collapse.  The
local-stiffness estimate motivates a plausible prompt-collapse scale, but does
not provide an upper bound on the mass that may ultimately enter the first
black hole.  At the same time, the large radius and finite hydrodynamic
response time of the SMDS leave open the possibility that outer layers respond
on a delayed timescale, remain bound, and form a quasi-star-like envelope or
fallback reservoir.  This possibility is strengthened by the extreme fallback
expected after black-hole formation: for
\(M_{\rm BH,0}\sim10^6\,M_\odot\), the initial fallback rate is many orders of
magnitude above the Eddington-regulated accretion rate, so the inner flow
should be photon trapped and capable of launching a radiation-pressure-mediated
shock or pressure wave into the surrounding stellar material.  As shown below,
the available prompt-collapse energy is substantial; however, by itself
it won't significantly inflate the entire envelope; the remaining energy can be supplied by the same
post-collapse accretion feedback that powers the quasi-star phase.

Determining whether the collapse separates into a prompt black hole plus a
delayed envelope, or instead proceeds through most of the star, requires a
nonlinear hydrodynamic or GR-hydrodynamic calculation, which is outside of the scope of this paper.  In the remainder of this work we use
$M_{BH,0}\sim 10^6 M_\odot$ as a fiducial parameter for estimating
the energetics and observational consequences of the SMDS QS scenario.

\subsection{Envelope binding energy and collapse energetics}\label{ssec:BindingEnergy}

The survival and subsequent inflation of the envelope depend on its \emph{net} binding energy, which is calculated using the relevant \texttt{MESA} profile as:
\begin{equation}\label{eq:EbindMESA}
    E_{\rm bind}^{\rm MESA}(m_{\rm cut})
    \equiv
    -\int_{m>m_{\rm cut}} e_{\rm tot}(m)\,\mathrm{d}m .
\end{equation}
The specific total energy may be written schematically as
\[
e_{\rm tot}(m)\simeq -\frac{Gm}{r}+u,
\]
where \(u\) is the specific internal energy, dominated here by radiation
(\(u\simeq 3P_{\rm rad}/\rho\)).  
Evaluating Eq.~\eqref{eq:EbindMESA} for the fiducial mass cut $m_{\rm cut}=M_{BH,0}\simeq10^6\,M_\odot$ gives
\begin{equation}\label{eq:BindingMcutE6}
    |E_{\rm bind}^{\rm MESA}| \simeq 7\times10^{56}\,{\rm erg}.
\end{equation}

As this is a key intermediary result of our manuscript, with important implications on the plausibility that collapsed SMDSs lead to late stage quasi-star like configurations, we validate it analytically in Appendix~\ref{app:binding_energy_validation}, where we derive useful closed form approximations for the binding energy of a radiation dominated shell using a modified form of the virial theorem for a truncated envelope. 

The part of the energy released during prompt collapse of the core which couples to the envelope can be approximated as a fraction of the characteristic gravitational energy liberated as $M_{\rm BH,0}$ contracts from its pre-collapse scale $R_{\rm core}$:
\begin{equation}
    E_{\rm collapse} \sim \alpha \frac{G M_{\rm BH,0}^2}{R_{\rm core}},
    \label{eq:Ecollapse}
\end{equation}
where $\alpha$ accounts for both the structural prefactor of the collapsing core and the fraction of the released energy that couples to the overlying envelope rather than being advected into the BH, lost to neutrinos, or retained in channels that do not do work on the envelope.  The \texttt{MESA} enclosed-mass profile gives $R_{\rm core}\simeq4\times10^3\,R_\odot$ for $M_{\rm BH,0}=10^6\,M_\odot$, implying
\begin{equation}
    E_{\rm collapse}
    \sim
    9.5\times10^{55}
    \left(\frac{\alpha}{0.1}\right)
    {\rm erg}.
    \label{eq:inflation_prompt_energy}
\end{equation}
Comparing this collapse-energy estimate to the corresponding \texttt{MESA} binding energy gives
\begin{equation}
    \frac{E_{\rm collapse}}{|E_{\rm bind}^{\rm MESA}|}
    \sim
    0.7\left(\frac{\alpha}{0.5}\right).
\end{equation}
Thus prompt collapse alone can account for the envelope binding energy only if the coupling parameter is large, $\alpha\sim0.7$, corresponding to highly efficient deposition of the collapse energy into the envelope.  More conservative values, $\alpha\simeq0.1$--0.3, provide an important initial perturbation but leave the envelope bound; in that case the remaining inflation energy must come from post-collapse accretion feedback, as in the standard SMS quasi-star pathway.

\section{Possible Formation of a Quasi-star-like Remnant from SMDS Collapse}\label{sec:qs_formation}

We now analyse the possible formation of a quasi-star-like remnant from the SMDS collapse product, emphasising the physical conditions that must be satisfied and contrasting with the SMS pathway.

\subsection{Quasi-star equilibrium conditions}\label{ssec:qs_equilibrium}

Following \citet{BegelmanRossiArmitage2008} and \citet{Begelman2010}, we define the quasi-star state as an envelope-regulated accretion configuration.  The central BH may accrete at a rate far above its own Eddington limit, but the luminosity transported through the extended envelope is limited by the Eddington luminosity of the \emph{total} gravitating mass,
\begin{equation}
    L_{\rm BH}
    =
    \varepsilon \dot{M}_{\rm BH} c^2
    \simeq
    L_{\rm Edd}(M_{\rm QS})
    =
    \frac{4\pi G M_{\rm QS} c}{\kappa_\mathrm{es}},
    \label{eq:qs_eddington_condition}
\end{equation}
where $M_{\rm QS}\simeq M_{\rm BH}+M_{\rm env}$ is the mass of the BH--envelope system, $\varepsilon$ is the radiative efficiency, and $\kappa_\mathrm{es}$ is the electron-scattering opacity.  For fully ionized primordial gas we take $\kappa_\mathrm{es}\simeq0.34\,\mathrm{cm}^2\,\mathrm{g}^{-1}$, appropriate for hydrogen mass fraction $X\simeq0.75$.  For the fiducial post-collapse system, $M_{\rm QS}\simeq2.6\times10^6\,M_\odot$, giving
\begin{equation}
    L_{\rm Edd}(M_{\rm QS})
    \simeq
  10^{11}\,L_\odot
    \simeq
    (3.9\text{--}4.3)\times10^{44}\,{\rm erg\,s^{-1}} .
    \label{eq:qs_edd_luminosity}
\end{equation}

The corresponding regulated BH accretion rate is
\begin{eqnarray}
    \dot{M}_{\rm BH,eq}
    &=& \frac{L_{\rm Edd}(M_{\rm QS})}{\varepsilon c^2}
    \nonumber\\
    &\simeq&
    6.8\times10^{-2}\,M_\odot\,\mathrm{yr}^{-1}
    \left( \frac{L_{\rm Edd}}{10^{11}\,L_\odot} \right)
    \left( \frac{0.1}{\varepsilon} \right)
    \label{eq:qs_mdot_eq}
\end{eqnarray}  
For the fiducial mass cut $M_{\rm BH,0}\simeq10^6\,M_\odot$, the remaining envelope mass is $M_{\rm env}\simeq1.6\times10^6\,M_\odot$, so the envelope reservoir could supply the regulated rate for
\begin{equation}
    t_{\rm res}
    \simeq
    2\times10^7\,{\rm yr}
    \left(\frac{M_{\rm env}}{1.6\times10^6\,M_\odot}\right)
    \left(\frac{0.07\,M_\odot\,{\rm yr}^{-1}}{\dot M_{\rm BH,eq}}\right),
    \label{eq:qs_reservoir_time}
\end{equation}

The required photospheric radius follows from the same envelope-regulated luminosity.  A quasi-star envelope evolves toward a cool, convective, radiation-pressure-supported structure whose photosphere is limited by a Hayashi-like minimum effective temperature, $T_{\rm eff,min}\sim3000$--$6000\,{\rm K}$ for primordial opacities~\citep{BegelmanRossiArmitage2008,VolonteriBegelman2010}.  Combining $L\simeq L_{\rm Edd}(M_{\rm QS})$ with the Stefan--Boltzmann law gives
\begin{equation}
    R_{\rm QS}
    \simeq
    4.2\times10^5\,R_\odot
    \left(\frac{L_{\rm Edd}}{10^{11}\,L_\odot}\right)^{1/2}
    \left(\frac{5000\,{\rm K}}{T_{\rm eff}}\right)^2 .
    \label{eq:qs_radius}
\end{equation}
For $T_{\rm eff}=6000$--$3000\,{\rm K}$ this corresponds to $R_{\rm QS}\simeq3\times10^5$--$1.2\times10^6\,R_\odot$.  Relative to the pre-collapse SMDS radius $R_\star\simeq1.7\times10^4\,R_\odot$, the required expansion factor is therefore
\begin{equation}
   f_R
    \equiv
    \frac{R_{\rm QS}}{R_\star}
    \simeq
    25
    \left(\frac{L_{\rm Edd}}{10^{11}\,L_\odot}\right)^{1/2}
    \left(\frac{5000\,{\rm K}}{T_{\rm eff}}\right)^2
    \left(\frac{1.7\times10^4\,R_\odot}{R_\star}\right).
    \label{eq:qs_inflation_factor}
\end{equation}
 A key issue, which we address next,  is whether the post-collapse remnant can plausibly supply enough energy to move the weakly bound SMDS envelope to this larger-radius branch without immediately unbinding it.

\subsection{Envelope inflation energetics}\label{ssec:envelope_inflation}

The energy scale for the required inflation is set by the initial binding energy of the envelope outside the prompt BH mass cut.  For the fiducial $m_{\rm cut}=M_{\rm BH,0}\simeq10^6\,M_\odot$, the direct \texttt{MESA} total-energy integral gives $|E_{\rm bind}^{\rm MESA}|
    \simeq
    7\times10^{56}\,{\rm erg}$ (see Eqns.~\ref{eq:EbindMESA}-\ref{eq:BindingMcutE6}).

We next estimate the energy needed for an envelope expansion by a factor $f_R\simeq25$. Assuming it remains virialized during its relaxation, the magnitude of its binding energy scales approximately as $|E|\propto R^{-1}$.  The energy required to move from the pre-collapse radius to the quasi-star radius is therefore of order
\begin{eqnarray}
   \Delta E_{\rm infl}
    &\sim&
    |E_{\rm bind}|
    \left(1-\frac{1}{f_R}\right)\\ \nonumber
    &&\simeq
    6.7\times10^{56}\,{\rm erg}
    \left(\frac{|E_{\rm bind}^{\rm MESA}|}{7\times10^{56}\,{\rm erg}}\right)
    \left(\frac{1-1/f_R}{0.96}\right).
    \label{eq:inflation_energy_required}
\end{eqnarray}
Thus the desired inflation requires an energy input comparable to, but slightly below, the original envelope binding energy. Motivated by this result, in what follows we will assume $\Delta E_{\rm infl}\sim |E_{\rm bind}|$.

Prompt collapse can provide a substantial first contribution, as discussed in Sec.~\ref{ssec:BindingEnergy}. The remaining energy can be supplied naturally by the same BH accretion feedback that defines the quasi-star phase.  If a fraction $f_{\rm dep}$ of the accretion luminosity is retained as envelope heat or mechanical work, then
\begin{equation}
    E_{\rm acc,dep}(t)
    =
    f_{\rm dep}
    \int_0^t L_{\rm BH}(t')\,{\rm d}t'
    \simeq
    f_{\rm dep} L_{\rm Edd}(M_{\rm QS}) t
    \label{eq:inflation_accretion_integral}
\end{equation}
after the system approaches envelope-regulated accretion.  The time to deposit one initial binding energy is
\begin{equation}
    t_{\rm bind}
    \sim
    \frac{|E_{\rm bind}^{\rm MESA}|}{f_{\rm dep} L_{\rm Edd}}
    \simeq
    6\times10^4\,f_{\rm dep}^{-1}\,{\rm yr}.
    \label{eq:inflation_binding_time}
\end{equation}
Equivalently, the accreted BH mass needed to supply this energy is
\begin{equation}
    \Delta M_{\rm BH}
    \sim
    \frac{|E_{\rm bind}^{\rm MESA}|}{f_{\rm dep}\varepsilon c^2}
    \simeq
    4\times10^3\,f_{\rm dep}^{-1}
    \left(\frac{0.1}{\varepsilon}\right)
    M_\odot .
    \label{eq:inflation_accreted_mass}
\end{equation}
Even for $f_{\rm dep}=0.1$, this corresponds to only $\sim4\times10^4\,M_\odot$, a few per cent of the bound envelope and far below the available reservoir.  If prompt collapse has already deposited part of the energy, the required accretion time and accreted mass are reduced further.

Following the initial inflation phase described above, the envelope reaches a quasi-star configuration with a photospheric temperature approaching the primordial Hayashi limit. This transition does not terminate accretion onto the embedded black hole. Rather, it marks the point at which the envelope ceases to behave primarily as an energy reservoir. During the inflation phase, a significant fraction of the accretion energy is converted into internal energy and mechanical work, causing the envelope to expand. Once the envelope has relaxed onto the Hayashi branch, however, additional accretion energy is transported outward through the envelope and radiated from the cool photosphere, while a fraction may also be carried away by radiation-driven winds or other forms of mass loss. The subsequent evolution is therefore regulated by the balance between accretion, radiative diffusion, convection, and mass loss, rather than by continued large-scale envelope inflation. As the black hole gradually grows at the expense of the envelope, the quasi-star evolves quasi-statically until the remaining envelope can no longer sustain hydrostatic equilibrium, at which point the quasi-star phase terminates and the residual gas is expected either to be expelled or rapidly accreted.


\subsection{Comparison with SMS-based quasi-stars}

The SMS quasi-star pathway passes through analogous stages, but under more restrictive conditions.
Table~\ref{tab:smds_sms_qs} summarises the key differences between the SMS and SMDS routes to quasi-star-like remnants.

\begin{deluxetable*}{llll}[t]
\tabletypesize{\scriptsize}
\tablewidth{0pt}
\tablecaption{Comparison of SMS and SMDS pathways to quasi-stars.\label{tab:smds_sms_qs}}
\tablehead{
\colhead{\parbox{0.16\textwidth}{Aspect}} & \colhead{\parbox{0.27\textwidth}{SMS $\rightarrow$ Quasi-star~\citep[e.g.][]{Begelman2010}}} & \colhead{\parbox{0.27\textwidth}{SMDS $\rightarrow$ Quasi-star (This Work)}} & \colhead{\parbox{0.22\textwidth}{Comment}}
}
\startdata
\parbox[t]{0.16\textwidth}{\raggedright Formation environment} & \parbox[t]{0.27\textwidth}{\raggedright Requires strong LW background to suppress H$_2$ in the host atomically cooled halo.} & \parbox[t]{0.27\textwidth}{\raggedright Forms in both H$_2$ and atomically cooled halos; does not rely on LW suppression for puffiness.} & \parbox[t]{0.22\textwidth}{\raggedright SMDS channel operates in a wider range of early environments.} \\
\parbox[t]{0.16\textwidth}{\raggedright Stellar-phase accretion} & \parbox[t]{0.27\textwidth}{\raggedright Needs $\dot{M}_\star \gtrsim 0.1$--$1\,M_\odot\,\mathrm{yr}^{-1}$ to keep a bloated, non-relaxed envelope.} & \parbox[t]{0.27\textwidth}{\raggedright DM heating allows much lower $\dot{M}_\star \sim 10^{-3}$--$10^{-2}\,M_\odot\,\mathrm{yr}^{-1}$.} & \parbox[t]{0.22\textwidth}{\raggedright SMDS does not require extreme inflow to stay extended.} \\
\parbox[t]{0.16\textwidth}{\raggedright Envelope structure} & \parbox[t]{0.27\textwidth}{\raggedright Hylotropic entropy profile required to keep envelope weakly bound.} & \parbox[t]{0.27\textwidth}{\raggedright Convective, but large radius from DM heating yields low binding energy.} & \parbox[t]{0.22\textwidth}{\raggedright SMDS envelope binding energy is smaller than that of a SMS of the same mass.} \\
\parbox[t]{0.16\textwidth}{\raggedright GR-instability threshold} & \parbox[t]{0.27\textwidth}{\raggedright Non-rotating SMS unstable at $\sim$ few $\times 10^5\,M_\odot$; rotation needed to reach $\sim 10^6\,M_\odot$.} & \parbox[t]{0.27\textwidth}{\raggedright Non-rotating SMDS reaches $\simeq 2.6\times10^6\,M_\odot$ at GR onset.} & \parbox[t]{0.22\textwidth}{\raggedright SMDS does not require rotational support to hit quasi-star masses.} \\
\parbox[t]{0.16\textwidth}{\raggedright Inflation energy source} & \parbox[t]{0.27\textwidth}{\raggedright Formation and early BH growth deposit energy into the envelope; sustained accretion powers the quasi-star.} & \parbox[t]{0.27\textwidth}{\raggedright Prompt collapse can perturb the envelope, while time-integrated accretion supplies the remaining inflation energy.} & \parbox[t]{0.22\textwidth}{\raggedright SMDS follows the standard quasi-star feedback logic, but starts from a larger, weakly bound envelope.} \\
\parbox[t]{0.16\textwidth}{\raggedright BH-to-envelope mass fraction} & \parbox[t]{0.27\textwidth}{\raggedright Canonical quasi-star solutions have $M_{\mathrm{BH},0}/M_\star \lesssim 10^{-2}$.\tablenotemark{\footnotesize{a}}} & \parbox[t]{0.27\textwidth}{\raggedright Nearly homologous collapse can lead to $M_\mathrm{BH,0}/M_\star \gtrsim 0.5$} & \parbox[t]{0.22\textwidth}{\raggedright SMDS starts in late-stage quasi-star regime.} \\
\parbox[t]{0.16\textwidth}{\raggedright LRD connection} & \parbox[t]{0.27\textwidth}{\raggedright Late Stage Quasi-Stars} & \parbox[t]{0.27\textwidth}{\raggedright Starts extended and weakly bound; easier to reach a cool quasi-star-like photosphere.} & \parbox[t]{0.22\textwidth}{\raggedright SMDS remnants can act as unresolved LRD-like luminous components.}
\enddata
\tablenotetext{a}{Recent saturated-convection quasi-star models show that hydrostatic envelopes can persist to large black-hole mass fractions after embedded BH growth, reaching $M_{\rm BH}\simeq0.6\,M_{\rm QS}$ in some solutions~\citep{Coughlin_Begelman_2024}.}
\end{deluxetable*}
The main message is that SMDSs relax or remove several environmental and structural fine-tunings required in the SMS channel while still satisfying the fundamental ingredients for a quasi-star-like phase: GR-induced core collapse, a massive bound envelope, and an energetically favourable regime for envelope inflation.

\section{Connection to JWST Little Red Dots}\label{sec:lrd_connection}

LRDs are characterised observationally by very red rest-optical colors, compact morphologies, and extreme luminosities at $z \gtrsim 7$ \citep[e.g.,][]{Kocevski2024,Wang2024,LiInayoshi2025,Taylor2025,VaidaFarber2026}.
SED fitting suggests that many LRDs can be modelled by cool photospheres with $T_\mathrm{eff} \sim 3000$--$6000\,$K and bolometric luminosities $L \sim 10^{11}$--$10^{12}\,L_\odot$, sometimes with additional dust reddening or line emission~\citep[e.g.][]{ronayne2025megaspectrophotometricsedfitting,
liu2025balmerbreakopticalcontinuum,santarelli2026evolutionarytracksspectralproperties}.
Such temperatures are reminiscent of Hayashi-limit envelopes in massive stars and quasi-star models.

In the quasi-star context, reproducing the red colors and high luminosities of LRDs requires a late-stage object: an envelope expanded to near the Hayashi limit while the interior BH has grown to at least $10\%$ of the total mass of the progenitor SMS~\citep{Begelman_2025}. As shown in the previous section, collapsed SMDSs naturally lead directly to this configuration. For a SMDS with $L\simeq L_{\rm Edd}(M_\star)\simeq 10^{11}\,L_\odot$, reaching $T_\mathrm{eff}\sim5000\,\mathrm{K}$ requires $R\simeq4.2\times10^5\,R_\odot$. This is only a factor $f_R\sim25$ above our fiducial $100$~GeV pre-collapse SMDS radius, compared with $f_R\sim10^2$--$10^3$ for conventional SMSs with radii $\sim10^2$--$10^3\,R_\odot$. The quasi-star photosphere would remain unresolved by JWST; any larger observed size must trace nebular emission, reprocessed radiation, or host-galaxy light.

Next we check that our SMDS quasi-star like configurations have sufficiently high H column densities in order to obscure the central BH, a key requirement of matching the LRDs JWST data. For the fiducial remnant with \(M_{\rm env}\simeq1.6\times10^6\,M_\odot\),
the inflated quasi-star radii implied by \(L\simeq L_{\rm Edd}(M_\star)\) and
\(T_{\rm eff}\simeq3000\)--\(6000\,{\rm K}\) correspond to an angle-averaged
surface mass density
\[
\Sigma_{\rm env}
=
\frac{M_{\rm env}}{4\pi R_{\rm QS}^2}
\simeq
3.6\times10^4\text{--}5.8\times10^5\,{\rm g\,cm^{-2}},
\]
or, equivalently, a hydrogen column
\[
N_{\rm H}
\simeq
\frac{X\Sigma_{\rm env}}{m_p}
\simeq
1.6\times10^{28}\text{--}2.6\times10^{29}\,{\rm cm^{-2}},
\qquad
X=0.75.
\]
Thus even after inflation to the LRD-like radius, the surviving envelope
remains deeply Compton-thick in total column and can efficiently obscure and
reprocess emission from the embedded BH.

\section{Conclusions}\label{sec:conclusions}

 We have examined the collapse of supermassive dark stars and, using a fiducial case of a SMDS powered by 100~GeV WIMPs, we argued that supermassive dark stars are natural progenitors of massive bound envelopes that may evolve into late-stage quasi-star-like objects. In turn, those match many of the observed properties of JWST's mysterious little red dots. 
 
Contrasting with the SMS-based quasi-star pathway, we find that SMDSs relax or remove several fine-tuned conditions: they do not require strong LW backgrounds, extreme gas inflow rates, or delicate rotational support to reach quasi-star-scale masses; their envelopes are naturally weakly bound due to DM heating being deposited throughout the star; and the heavy BH seeds produced by SMDS collapse lead to large (e.g. $\gtrsim 0.5$) BH-to-envelope mass fractions, which are much larger than those for canonical SMS quasi-stars, yet compatible with recent saturated-convection models~\citep{Coughlin_Begelman_2024}.

The SMDS to quasi-star like objects channel thus provides a physically motivated route to unresolved, LRD-like luminous components and massive BH seeds in the early universe.
In future work we will perform a  radial pulsation stability analysis, couple GR collapse simulations of SMDSs to post-collapse envelope evolution (e.g., via \texttt{GR1D} or similar codes), and perform radiative transfer calculations for the resulting atmospheres to compare directly with JWST LRD spectra.

\begin{acknowledgments}
 C.I. acknowledges funding from Colgate University
via the Research Council (Grant No. 821028) and the Picker 
Interdisciplinary Science Institute (Grant No. 826837). C.I. thanks Sohan Ghodla for sharing the 100 GeV \texttt{MESA} SMDS profiles we used in \cite{FreeseIlie2025}, and which we adopted here as a fiducial model. We furthermore acknowledge the use of Colgate's Turing Supercomputer 
(Partially supported by NSF grant OAC-2346664).
\end{acknowledgments}

\appendix

\section{Analytic Validation of the Envelope Binding Energy}
\label{app:binding_energy_validation}

In this Appendix we validate the envelope binding energies used in the main text by comparing the direct \texttt{MESA} energy integral against an independent estimate based on the virial theorem with the appropriate surface-pressure term.  We then specialize this estimate to an $n=3$ polytrope, appropriate for a radiation-pressure dominated supermassive star, and derive a useful expansion in the prompt-collapse mass cut.

\subsection{Direct \texttt{MESA} estimate}

For a mass cut $m_{\rm cut}$, identified with the prompt black-hole seed mass $M_{\rm BH,0}$, we define the binding energy of the remaining envelope as the negative of the total specific energy integrated over the exterior layers:
\begin{equation}
E_{\rm env}(>m_{\rm cut})
=
\int_{m_{\rm cut}}^{M_\star}
e_{\rm tot}(m)\,dm,
\end{equation}
with
\begin{equation}
E_{\rm bind}^{\rm MESA}
=
\max\left[0,-E_{\rm env}(>m_{\rm cut})\right].
\end{equation}
Here $e_{\rm tot}$ is the \texttt{MESA} total specific energy, including internal and gravitational contributions.  This is the primary binding-energy estimate used in the main text.

\subsection{Modified virial theorem for a truncated envelope}

For an envelope extending from an inner boundary $r_{\rm cut}$ to the stellar surface $R_\star$, hydrostatic equilibrium gives
\begin{equation}
{dP\over dr}
=
-{Gm(r)\rho\over r^2}.
\end{equation}
Multiplying by $4\pi r^3dr$ and integrating over the envelope gives
\begin{equation}
\left[4\pi r^3P\right]_{r_{\rm cut}}^{R_\star}
-
3\int_{r_{\rm cut}}^{R_\star}P\,dV
=
W_{\rm env},
\end{equation}
where $W_{\rm env}<0$ is the gravitational potential energy of the envelope, including the interaction with the interior mass.  Therefore
\begin{equation}
W_{\rm env}
+
3\int_{r_{\rm cut}}^{R_\star}P\,dV
-
4\pi R_\star^3P(R_\star)
+
4\pi r_{\rm cut}^3P(r_{\rm cut})
=
0.
\end{equation}
Since $P(R_\star)\simeq0$, this becomes
\begin{equation}
W_{\rm env}
+
3\int_{r_{\rm cut}}^{R_\star}P\,dV
+
4\pi r_{\rm cut}^3P(r_{\rm cut})
\simeq
0.
\end{equation}
For a radiation-pressure dominated envelope, $u\simeq3P$, so that
\begin{equation}
U_{\rm env}
\simeq
3\int_{r_{\rm cut}}^{R_\star}P\,dV.
\end{equation}
The shell virial theorem is then
\begin{equation}
W_{\rm env}
+
U_{\rm env}
+
4\pi r_{\rm cut}^3P(r_{\rm cut})
\simeq
0.
\label{eq:app_shell_virial}
\end{equation}
Thus the total energy of the exterior envelope is
\begin{equation}
E_{\rm env}
=
W_{\rm env}+U_{\rm env}
\simeq
-4\pi r_{\rm cut}^3P(r_{\rm cut}),
\end{equation}
and the absolute value of the binding energy is
\begin{equation}
\boxed{
E_{\rm bind,env}
\simeq
4\pi r_{\rm cut}^3P(r_{\rm cut}).
}
\label{eq:app_surface_binding}
\end{equation}
This expression shows that the binding energy of a radiation-dominated truncated envelope is controlled mainly by the pressure work at the inner boundary.

\subsection{\texorpdfstring{$n=3$}{n=3} polytropic estimate}

For an $n=3$ Lane-Emden polytrope,
\begin{equation}
r=a\xi,
\qquad
\rho=\rho_c\theta^3,
\qquad
P=P_c\theta^4,
\end{equation}
and
\begin{equation}
m(\xi)=4\pi a^3\rho_c\mu(\xi),
\qquad
\mu(\xi)\equiv-\xi^2\theta'(\xi).
\end{equation}
The surface is at
\begin{equation}
\xi_1=6.89685,
\qquad
\mu_1\equiv\mu(\xi_1)=2.01824.
\end{equation}
Define the dimensionless mass cut
\begin{equation}
q\equiv {m_{\rm cut}\over M_\star}.
\end{equation}
The corresponding Lane-Emden coordinate $\xi_{\rm cut}$ is determined by
\begin{equation}
q
=
{\mu(\xi_{\rm cut})\over\mu_1}
=
{-\xi_{\rm cut}^2\theta'(\xi_{\rm cut})\over\mu_1}.
\label{eq:app_q_xi}
\end{equation}
The central pressure of an $n=3$ polytrope can be written in terms of the stellar mass and radius as
\begin{equation}
P_c
=
{G M_\star^2\over R_\star^4}
{\xi_1^4\over16\pi\mu_1^2}
\simeq
11.05\,{G M_\star^2\over R_\star^4}.
\end{equation}
Using $r_{\rm cut}=R_\star\xi_{\rm cut}/\xi_1$ and $P_{\rm cut}=P_c\theta(\xi_{\rm cut})^4$, Equation~(\ref{eq:app_surface_binding}) gives
\begin{equation}
E_{\rm bind,env}^{n=3}(q)
=
{G M_\star^2\over R_\star}
F_3(q),
\end{equation}
where
\begin{equation}
\boxed{
F_3(q)
=
{\xi_1\over4\mu_1^2}
\xi_{\rm cut}^3\theta(\xi_{\rm cut})^4,
}
\label{eq:app_F3_exact}
\end{equation}
with $\xi_{\rm cut}$ determined implicitly by Equation~(\ref{eq:app_q_xi}).

\subsection{Explicit expansion in the mass cut}

Although Equation~(\ref{eq:app_F3_exact}) is compact, its dependence on $m_{\rm cut}$ is implicit.  For the mass cuts relevant here, it is useful to expand in powers of $q$. Keeping only the first two sub-leading terms, which as we show later is sufficient to estimate the binding energy to accuracy of $1\%$ or less,  the result of the expansion is
\begin{equation}
\boxed{
F_3(q)
\simeq
{3\xi_1\over4\mu_1}q
\left[
1
-
{11\over30}(3\mu_1 q)^{2/3}
+
{13\over840}(3\mu_1 q)^{4/3}
\right].
}
\label{eq:app_F3_expansion}
\end{equation}
Since
\begin{equation}
{3\xi_1\over4\mu_1}
\simeq
2.56,
\end{equation}
the binding energy may be written as
\begin{equation}
\boxed{
E_{\rm bind,env}(>m_{\rm cut})
\simeq
2.56\,{G M_\star m_{\rm cut}\over R_\star}
\left[
1
-
{11\over30}
\left(
3\mu_1{m_{\rm cut}\over M_\star}
\right)^{2/3}
+
{13\over840}
\left(
3\mu_1{m_{\rm cut}\over M_\star}
\right)^{4/3}
\right].
}
\label{eq:app_binding_explicit}
\end{equation}
The leading term is linear in $m_{\rm cut}$.  Physically, near the center of an $n=3$ polytrope the pressure varies slowly, $P_{\rm cut}\simeq P_c$, while $r_{\rm cut}^3\propto m_{\rm cut}$.  The correction terms in Equation~(\ref{eq:app_binding_explicit}) account for the decrease of pressure as the cut moves outward.

\subsection{Numerical comparison}

For the fiducial \texttt{MESA} profile used in the main text,
\begin{equation}
M_\star\simeq2.93\times10^6\,M_\odot,
\qquad
R_\star\simeq1.67\times10^4\,R_\odot,
\end{equation}
so that
\begin{equation}
{G M_\star^2\over R_\star}
\simeq
1.95\times10^{57}\,{\rm erg}.
\end{equation}
Table~\ref{tab:app_binding_validation} compares the direct \texttt{MESA} binding-energy integral with the modified-virial estimate evaluated from the \texttt{MESA} pressure and radius at the cut, and with the $n=3$ polytropic estimate.

\begin{table}[h]
\centering
\caption{Validation of the envelope binding energy.}
\label{tab:app_binding_validation}
\begin{tabular}{c c c c c c}
\hline
$m_{\rm cut}$ 
& $q$ 
& $E_{\rm bind}^{\rm MESA}$ 
& $4\pi r_{\rm cut}^3P_{\rm cut}$ 
& $E_{\rm bind}^{n=3}$ 
& $E_{\rm bind}^{\rm exp}$ \\
$(M_\odot)$ 
& 
& $(\mathrm{erg})$ 
& $(\mathrm{erg})$ 
& $(\mathrm{erg})$ 
& $(\mathrm{erg})$ \\
\hline
$10^5$ 
& $0.0341$ 
& $1.4\times10^{56}$ 
& $1.40\times10^{56}$ 
& $1.49\times10^{56}$ 
& $1.49\times10^{56}$ \\
$10^6$ 
& $0.341$ 
& $7.1\times10^{56}$ 
& $7.2\times10^{56}$ 
& $7.66\times10^{56}$ 
& $7.60\times10^{56}$ \\
\hline
\end{tabular}
\end{table}

For $m_{\rm cut}=10^5\,M_\odot$, the expansion in Equation~(\ref{eq:app_F3_expansion}) differs from the full Lane-Emden value of $F_3$ by only $0.0035\%$.  For $m_{\rm cut}=10^6\,M_\odot$, the difference is still only $0.8\%$.  Thus the explicit expansion is accurate for both mass cuts considered here.  The leading-order linear approximation alone is not sufficient for the larger mass cut, but the expression including the $q^{2/3}$ and $q^{4/3}$ corrections remains accurate at the percent level.

The agreement between the direct \texttt{MESA} energy integral, the surface-pressure form of the modified virial theorem, and the independent $n=3$ polytropic estimate validates the binding-energy scale used in the main text.  The result also clarifies the physical origin of the binding energy: for a radiation-pressure dominated truncated envelope, the dominant contribution is the pressure work at the inner boundary of the envelope, $4\pi r_{\rm cut}^3P(r_{\rm cut})$, rather than the virial energy of an isolated self-gravitating shell.

\bibliographystyle{aasjournal}
\bibliography{refs}

\end{document}